\documentclass[twocolumn]{article} %

\usepackage[colorlinks=true,     linkcolor=blue, citecolor=blue, filecolor=blue, urlcolor=blue]{hyperref}
\usepackage{gensymb}
\usepackage{xcolor}
\usepackage{graphicx}
\usepackage{natbib}
\usepackage{txfonts}

\newcommand{\markme}[1]{#1}

\usepackage{tabularx}
\usepackage{gensymb,ulem}
\newcommand{\B}{B}
\newcommand{\Tp}{T_{\textrm{p}}}
\newcommand{\Sp}{S_{\textrm{p}}}
\newcommand{\np}{n_{\textrm{p}}}

\newcommand{\vp}{v_{\textrm{p}}}
\newcommand{\va}{v_{\textrm{A}}}
\newcommand{\ac}{a_{\textrm{col,p-p}}}
\newcommand{\nO}{n_{\textrm{O}^{7+}}/n_{\textrm{O}^{6+}}}
\newcommand{\lnO}{\log{n_{\textrm{O}^{7+}}/n_{\textrm{O}^{6+}}}}

\newcommand{\Osix}{\textrm{O}^{6+}}
\newcommand{\Oseven}{\textrm{O}^{7+}}

\newcommand{\Te}{T_{\textrm{exp}}}
\newcommand{\qFe}{\tilde{q}_{\textrm{Fe}}}

\usepackage{caption}
\begin{document} 
\title{Challenges in identifying coronal hole wind}
\author{Verena Heidrich-Meisner$^1$, Sophie Teichmann$^2$, Lars Berger$^1$, and Robert F. Wimmer-Schweingruber$^1$\\ $^1$: Christian Albrechts University at Kiel, Germany, $^2$: \markme{TUD Dresden University of Technology, Germany}}


\date{}
   \maketitle

\abstract{
The solar wind is frequently categorized based on its respective solar source region. Solar wind originating in coronal holes is consequently called coronal hole wind. Two well-established coronal hole wind categorizations, the charge state composition-based \citet{zhao2009global} scheme and the proton-plasma based \citet{xu2014new}, identify a very different fraction of solar wind in the data from the Advanced Composition Explorer (ACE) as coronal hole wind during the solar activity minimum at the end of Solar cycle 24.
}
{
  Here, we investigate possible explanations why \citet{zhao2009global} sees (almost) only coronal hole wind in 2009 while \citet{xu2014new} identifies almost no coronal hole wind at the same time. The high fraction of \citet{zhao2009global} coronal hole wind in 2009 has also been addressed in  \citet{wang2019observations}.
}
{
  We compare the properties of the respective coronal hole wind types and their changes with the solar activity cycle, namely 2001-2010. As a comparison reference, we include coronal hole wind identified by an unsupervised machine learning approach, $k$-means, in our analysis.
}
{
 Because the ratio of the $\Oseven$ to $\Osix$ densities drops systematically for all solar wind during the solar activity minimum which cannot be captured by the fixed threshold on the O charge ratio suggested in \citet{zhao2009global}, we find that the \citet{zhao2009global} scheme likely misidentifies slow solar wind as coronal hole wind during the solar activity minimum. The $k$-means coronal hole wind agrees with the very low fraction of coronal hole wind observed by the \citet{xu2014new} classification. In addition, the $k$-means classification considered here includes two types of coronal hole wind, wherein the first is dominant during the solar activity maximum and exhibits comparatively higher  O and Fe charge states, whereas the second is dominant during the solar activity minimum and features lower O and Fe charge states. A low fraction of coronal hole wind from low-latitude coronal holes observed by ACE in 2009 is plausible since during this time period a very small number of low-latitude coronal holes is observed \citep{fujiki2016long}. \cite{xu2014new} and $7$-means also rely on fixed decision boundaries, wherein the \citet{xu2014new} decision boundary for coronal hole wind appears to be better adapted for solar activity minimum than maximum conditions.}
{
  The results imply that origin-oriented solar wind classification needs to be revisited and suggest that explicitly including the phase of the solar activity cycle can be expected to improve solar wind classification. 
}


%

\section{Introduction}
\begin{figure*}
  \includegraphics[width=\textwidth]{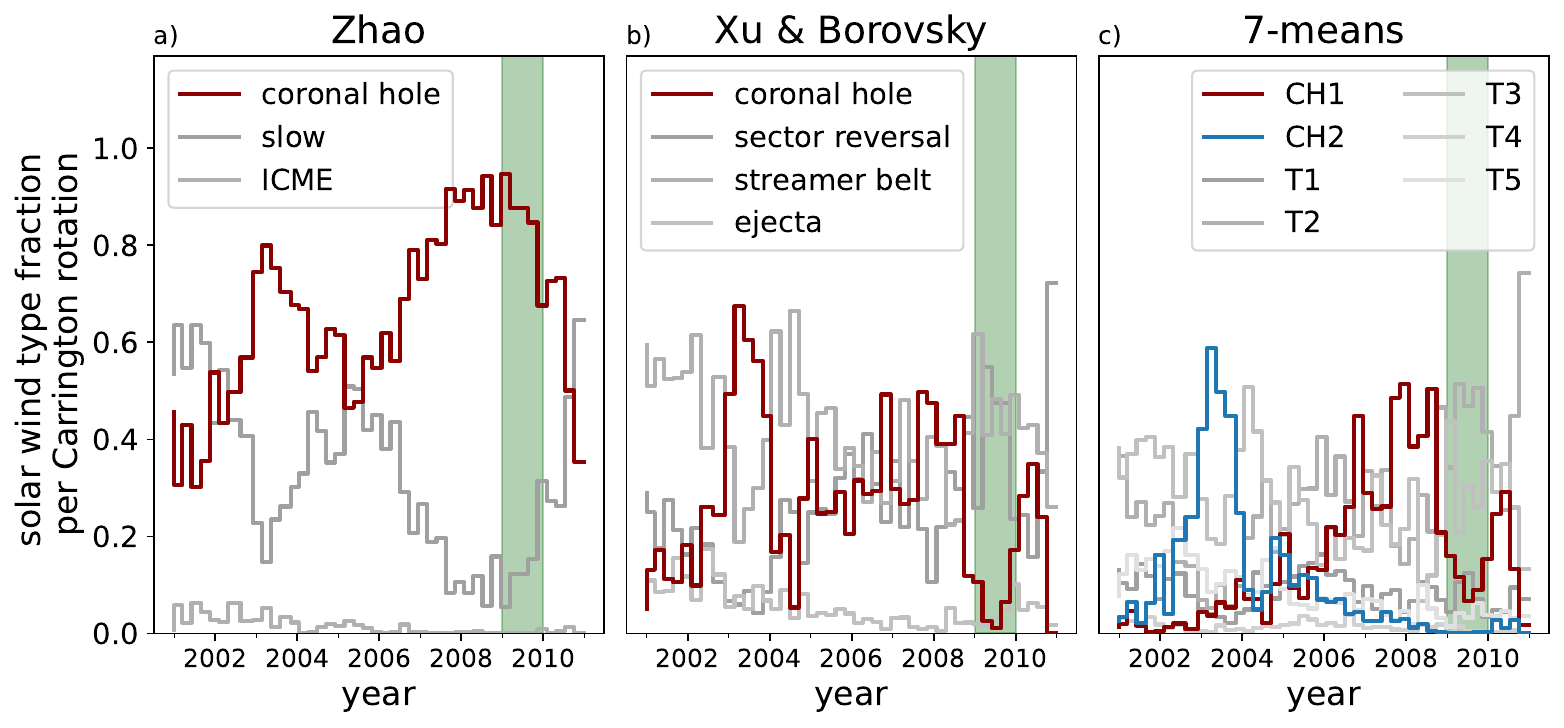}
  \caption{\label{fig:fractions} Fractions of solar wind types from three solar wind classifications per three Carrington rotations. From left to right: a)  \citet{zhao2009global}, b) \citet{xu2014new}, and c) $7$-means clustering. Green shading highlights the year 2009. All non-coronal hole wind types (Panel a): slow and ICME, Panel b): sector-reversal, helmet streamer, and ejecta, Panel c): the other five types identified by $7$-means which are here numbered as types T1, \dots, T5) are shown in different shades of grey to keep the focus on the coronal hole wind types we are interested in here.}
\end{figure*}

The solar wind, the plasma that is continuously emitted by the Sun, has been investigated for several decades \citep{parker1965dynamical,marsch2006kinetic,cranmer2009coronal,verscharen2019multi,vidotto2021evolution}.
The properties of the solar wind depend on the conditions in the corona at its solar source region and are modified by transport effects. The best-understood flavour of solar wind is called coronal hole wind since its source region has been identified as coronal holes \citep{hundhausen1968state,krieger1973coronal,Tu2005,cranmer2009coronal}. Compared to the so-called slow solar wind, coronal hole wind is considered better understood \citep{cranmer2009coronal}. Recently, the release mechanism of coronal hole wind has been linked to interchange reconnection \citep{bale2023interchange} and picoflares \citep{chitta2023picoflare}. \citet{zhao2014polar} has shown that the properties of coronal hole wind from polar coronal holes differ from those from  equatorial coronal holes. This effect is correlated to the footpoint field strength in the respective coronal hole \citep{wang2019observations,d2019slow,bale2019highly,stansby2019diagnosing,d2021alfvenic,wang2024coronal}.

The coronal hole wind is a collisionless plasma and, therefore, non-thermal features play an important role in its properties. Waves are frequently observed in coronal hole wind  \citep{marsch1982wave,d2015origin}. Wave-particle interactions lead to various interesting features in coronal hole wind, for example, a proton beam \citep{marsch1982solar}, differential streaming between protons and heavier ions \citet{marsch1981pronounced,berger2011systematic}, and temperature anisotropies \citep{marsch2006limits,kasper2003solar,huang2020proton}. Due to the variety of non-thermal features coronal hole wind is an interesting object of study (see, for example, the role of the Alfv{\'e}incity parameter in \citet{d2019slow,bale2019highly,stansby2019diagnosing,d2021alfvenic}) even though its solar source region is comparatively well understood.

To study solar wind from different solar sources separately, solar wind classification is often employed. The most prominent solar wind classification approaches focus on determining the respective solar source region \citep{zhao2009global, xu2014new, camporeale2017classification}. More recently, unsupervised machine learning approaches have been applied to solar wind categorization \citep{bloch2020data, heidrich2018solar, amaya2020visualizing}. In these data-driven methods, more solar wind types are identified that represent a mixture of source and transport-affected properties. In the following, we utilize a $k$-means solar wind classification as a data-driven reference classification. $k$-means is chosen here, since -- similar to \citet{zhao2009global} and \citet{xu2014new} --  $k$-means relies on fixed decision boundaries. 

\citet{neugebauer2016comparison} showed that although established solar wind categorizations agree on the basic types of solar wind, they disagree on how exactly the boundaries are defined. To the best of our knowledge, a ground truth for solar wind categorization is still not available. 

Here, we focus on the coronal hole wind identified by three different solar wind categorizations, the charge-state composition based \citet{zhao2009global} scheme, the proton-plasma property based \citet{xu2014new} scheme, and an unsupervised machine learning approach, $k$-means similar to \citet{bloch2020data} and \citet{heidrich2018solar}. Based on \citet{heidrich2018solar}, $k=7$ was chosen as a reference case for this study. Our study is motivated by the very different fractions of coronal hole wind identified by these three approaches. This is illustrated in Fig.~\ref{fig:fractions} which shows for each of the three categorizations the varying fractions of solar wind types observed by the Advanced Composition Explorer (ACE) in 2001-2010. Here, all non-coronal hole wind types are shown in different shades of grey and coronal hole wind in dark red and blue, respectively. The most striking difference (highlighted with green shading) occurs in 2009 during the solar activity minimum. Here, the composition-based \citet{zhao2009global} scheme identifies almost all solar wind as coronal hole wind, while the \citet{xu2014new} scheme detects almost no coronal hole wind in the same time period. The unsupervised machine learning approach, 7-means, includes two types of coronal hole wind and both of them make up only a small fraction of the solar wind observed in 2009.  This figure is discussed in more detail in Sect.~\ref{sec:comp} after an overview of the underlying data set in Sect.~\ref{sec:data} and the three solar wind categorizations of interest in Sect.~\ref{sec:class}. The high fraction of coronal hole wind identified by the \citet{zhao2009global} approach is already discussed in detail in \citet{wang2016role,wang2019observations} which shows that the O charge state composition is related to the footpoint magnetic field strength at the likely source region. Here, we focus on the comparison with the other coronal hole wind characterizations and on explanations which solar wind parameters contribute to the identification of coronal hole wind.

\section{Data and methods}
In this section, we give a brief overview of the data and methods employed in this study.
\subsection{Data}\label{sec:data}

Our aim is to compare coronal hole wind types identified by different solar wind categorizations that, in part, rely on charge state composition. Therefore,  we utilize data observed by three instruments on ACE: the magnetic field strength $B$ from the magnetometer (MAG, \citet{smith1998ace}), the proton speed $\vp$, proton density $\np$, and proton temperature $\Tp$ from the  Solar Wind Electron Proton Alpha Monitor (SWEPAM,\citet{mccomas1998unusual}), and the heavy ion charge-state composition (ratio of the $\Oseven$ density to the $\Osix$ density $\nO$ and the average Fe charge state $\qFe$ based on the Fe$^{8+}$ - Fe$^{13+}$ Fe charge states) from the Solar Wind Ion Composition Spectrometer (SWICS, \citet{gloeckler-etal-1998}). To ensure a minimum of statistics, the average Fe charge is only considered if at least ten Fe ions were observed. We verified that the results do not change systematically if this threshold is modified. We chose a small threshold since the larger the threshold is this introduces a stronger  selection bias towards denser solar wind. This bias is expected to reduce the number of available data points in particular for coronal hole wind from polar coronal holes during solar activity minimum conditions. The small threshold thus represents a compromise between sufficient statistics for the determination of the charge state distributions and the number of available data points to which we can apply our analysis. The MAG and SWEPAM data is binned to the native 12-min time resolution of SWICS.  To avoid data gaps in the SWEPAM data, the merged SWEPAM-SWICS data set is utilized. The MAG and SWEPAM data sets are taken from the ACE Science Center. 
Interplanetary Coronal Mass ejections (ICMEs) are independently taken from published ICME lists \citep{jian2006properties,jian2011comparing,cane2003interplanetary,richardson2010near}. 
As a derived parameter, we consider the proton-proton collisional age (also called Coulomb number in \citet{kasper2012evolution}). As discussed in \citet{heidrich2020proton} the proton-proton collisional age is well-suited to reconstruct the \cite{xu2014new} solar wind types.  The relevant part of the data set is available at \citet{Berger2023-gd}. The heavy ion composition is derived from the pulse-height-analysis (PHA) words as described in \citet{berger2008velocity}.

\begin{figure}
  \includegraphics[width=\columnwidth]{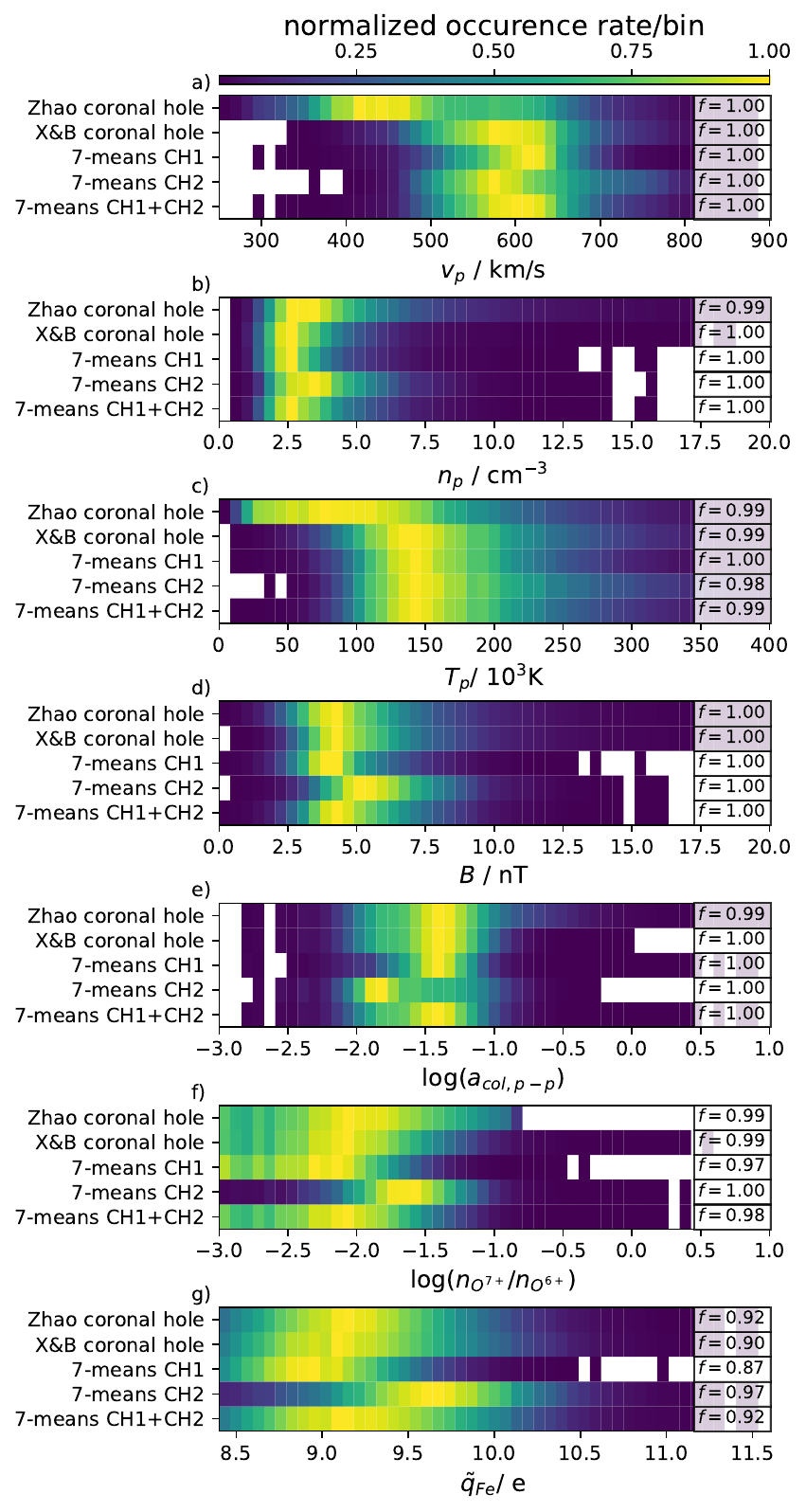}
  \caption{\label{fig:hist1d} 1-dimensional histograms of solar wind properties for five different coronal hole wind characterizations: \cite{zhao2009global} coronal hole, referred to in this and the following figures as \textit{Zhao coronal hole}), \citet{xu2014new} coronal hole, referred to in this and the following figures as \textit{X\&B coronal hole}, 7-means CH1, 7-means CH2, and a combination of the two 7-means coronal hole wind types which is referred to as 7-means CH1+CH2. Solar wind properties from top (a) to bottom (g): proton speed $v_p$, proton density $n_p$, proton temperature $T_p$, magnetic field strength $B$, proton-proton collisional age $\log(\ac)$, O charge state ratio $\log(\nO)$, average Fe charge state $\qFe$. The histograms in the Panels e) and f) are weighted by the logarithmic bin size. All 1-d histograms (with 50 bins each) are normalized to  their respective individual maximum. The fraction $f$ of included data points in each histogram is given as an inset. }
\end{figure}

\subsection{Feature explanations}\label{sec:shap}
In Sect.~\ref{sec:comp}, we employ a machine learning tool to illustrate and analyze the differences between the four coronal hole wind types of interest. In recent years, the concept of explainability has gained attention in the machine learning community (see, for example, \citet{solorio2020review,kumar2014feature,alelyani2018feature}). In contrast to black box results, explainability in this context refers to tools that provide interpretable explanations for the results of a machine learning method.
Some of these tools, in particular, posthoc explainability concepts like SHAP (SHapley additive explanations \citep{lundberg2017unified}) values, can also be applied to non-machine learning approaches. SHAP values are based on Shapeley values \citep{shapley1953value} that were originally designed for ranking players in collaborative games with teams of varying sizes. \citet{lundberg2017unified} proposed to apply Shapely values to explainability by interpreting the input parameters, in our case the solar wind properties ($\vp$, $\np$, $\Tp$, $B$, $\ac$, $\nO$, $\qFe$), as individual players in such a collaborative game. Teams of players then represent combinations of several input parameters. SHAP values can only be applied to regression or binary classification tasks directly. Here, we apply this concept to binary classification. Therefore, we treat the identification of coronal hole wind by each of the classifications as a binary classification task wherein a value of $1$ represents the respective coronal hole wind type and a value of $0$ represents any other solar wind type.

SHAP values then quantify how important each player, that is, each solar wind parameter, is to ``win'' the game. In this context, ``winning'' \footnote{In this context ``winning'' or ``loosing'' is not associated with any preference which result would be better.} corresponds to identifying data points as coronal hole wind.  High (positive) SHAP values then indicate  solar wind properties that are important to identify data points as coronal hole wind, whereas negative SHAP values with high absolute values indicate solar wind parameters that are important to identify non-coronal hole wind. SHAP values close to zero always represent unimportant parameters. In this study, we focus on positive SHAP values. SHAP values are computed for each data point separately. An overall importance of each feature can be obtained by summing all respective SHAP values. Here, we are interested in the change of the importance of respective solar wind parameters over time, therefore, we compute SHAP values Carrington-rotation wise. In addition to assessing the importance of solar wind parameters above or below the respective median value, the SHAP values for solar wind parameters above the median and below the median are represented separately. SHAP values assess the importance of single parameters and provide no information on the importance of tuples of input parameters.



\section{Coronal hole wind classification approaches}
\label{sec:class}
\begin{figure*}\begin{center}
  \begin{minipage}{0.9\columnwidth}
    \includegraphics[width=0.9\columnwidth]{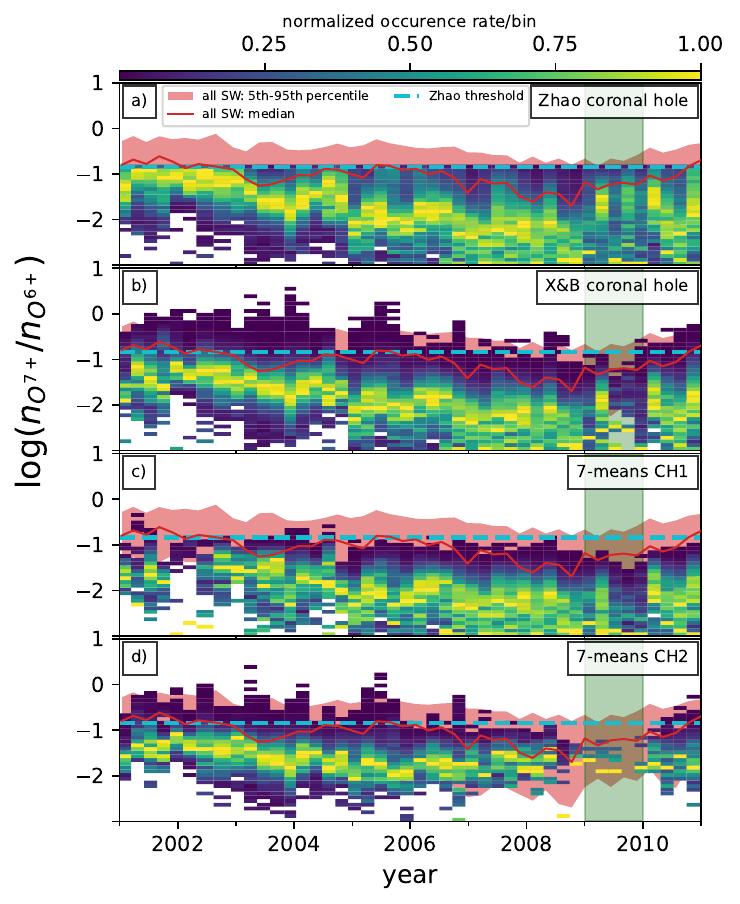}

    \includegraphics[width=0.9\columnwidth]{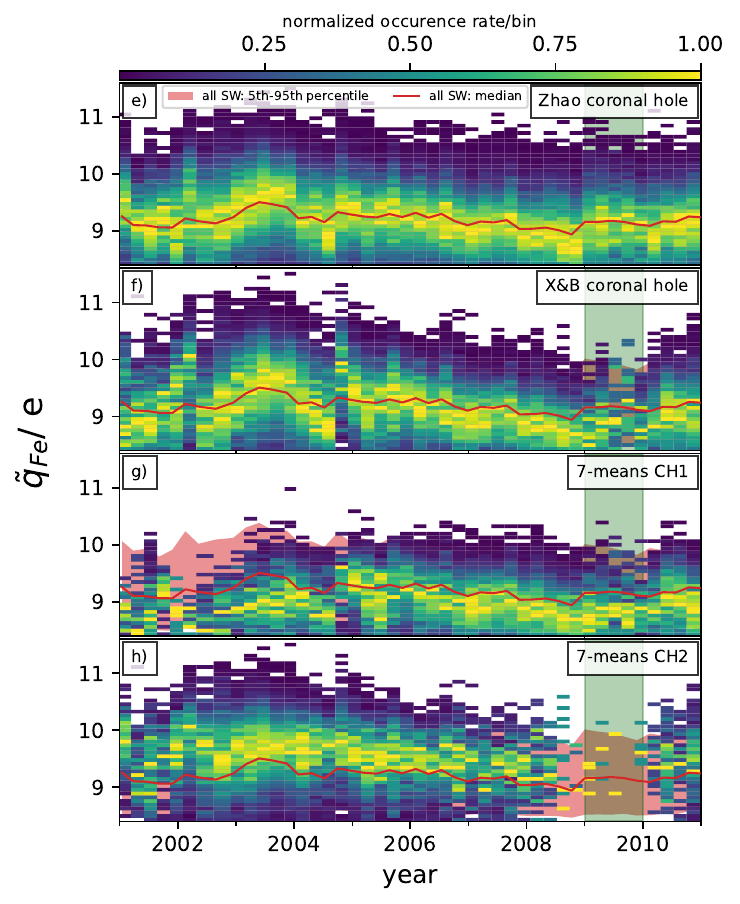}

  \end{minipage}
  \begin{minipage}{0.9\columnwidth}
  \includegraphics[width=0.9\columnwidth]{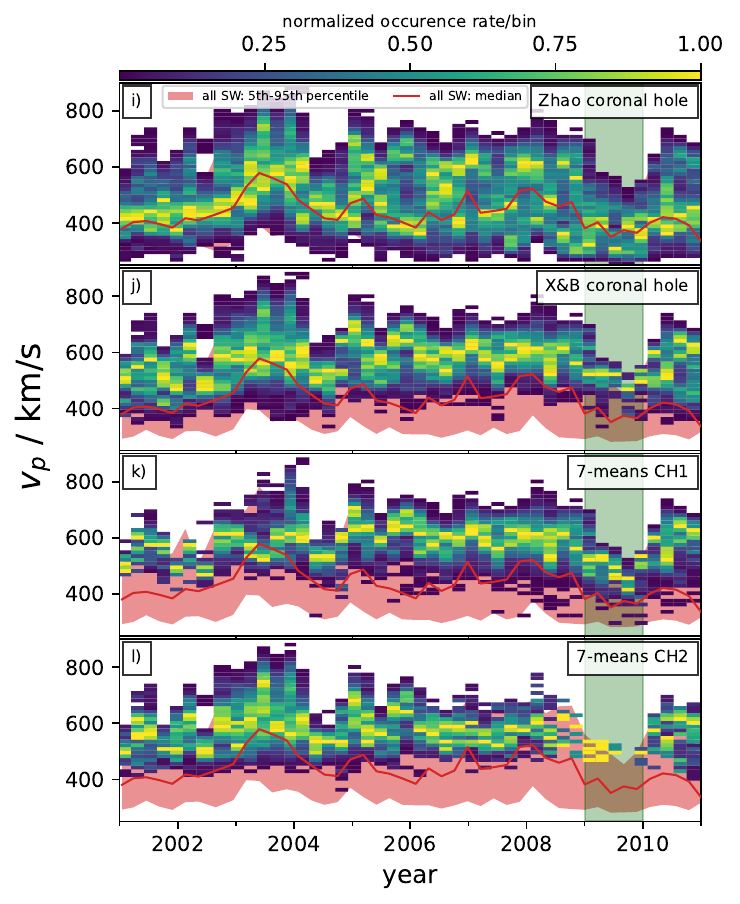}

    \includegraphics[width=0.9\columnwidth]{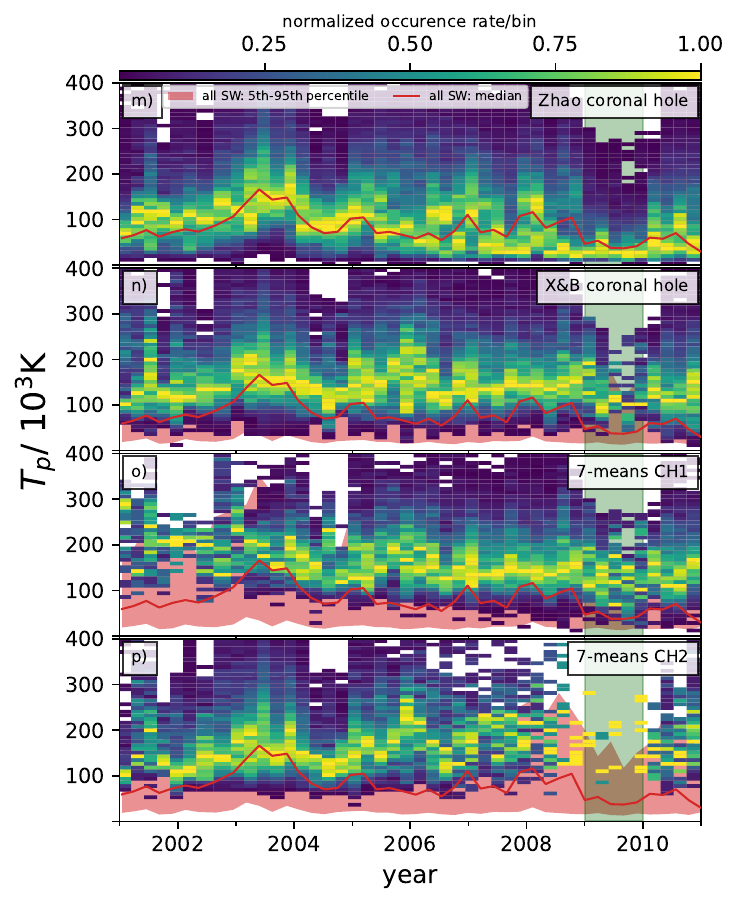}
\end{minipage}
  \end{center}
  \caption{\label{fig:cr} 1-dimensional histograms of solar wind properties over time. Each 1-d histogram represents three consecutive Carrington rotations.  $\nO$ in the top left (Panels a)-d)), $\qFe$ in the top right (Panels e)-h)), $\vp$ in the bottom left (Panels i)-l)), $\Tp$ in the bottom right (Panels m)-p)). In each panel, each column is normalized to its own maximum.  In each panel, the red line indicates the respective median for all solar wind observations and a red shading estimates the range of all solar wind observations in each Carrington rotation from the $5$th to $95$th percentile. Each subfigure shows the respective histograms for each of the four coronal hole wind types of interest. From top to bottom: \cite{zhao2009global} coronal hole, \citet{xu2014new} coronal hole, 7-means CH1, and 7-means CH2. In Panels a)-d), the decision threshold from \citet{zhao2009global} is indicated with a dashed cyan horizontal line. Green shading in all panels highlights the year 2009.}
\end{figure*}

In this study, we consider three solar wind classification approaches: \begin{enumerate}
\item The charge-state composition based  \citet{zhao2009global} scheme distinguishes between coronal hole wind and slow solar wind with a fixed threshold on the O charge state ratio $\nO$ and identifies all solar wind with $\nO<0.145$ as coronal hole wind. ICMEs are here distinguished from slow solar wind based on the proton speed $\vp$. 
\item The proton-plasma based \citet{xu2014new} scheme defines fixed decision boundaries based on the Alfv{\'e}n speed $\va$, the specific proton entropy $\Sp$, and a ratio between the observed proton temperature $\Tp$ and an expected proton-speed dependent temperature $\Te$ taken from \citet{elliott2005improved}. Here, coronal hole wind is identified by high Alfv{\'e}n speeds and high specific proton entropy. In most figures \citet{xu2014new} is abbreviated as X\&B.
\item As a reference comparison we include a $k$-means \citep{lloyd1982least} classification with $k=7$ similar to \citet{heidrich2018solar} and \citet{bloch2020data} based on proton speed $\vp$, proton density $\np$, proton temperature $\Tp$, magnetic field strength $B$, the proton-proton collisional age $\ac$, the O charge-state ratio $\nO$ and the average Fe charge state $\qFe$. Thus, the $7$-means classification combines proton plasma properties with charge state information. $k$-means was chosen here as a purely data-driven reference method that -- similar to \citet{zhao2009global} and \citet{xu2014new} -- relies on fixed decision boundaries. The $7$-means classification was trained on the ACE data from 2001-2010. However, ICMEs are not explicitly considered in $7$-means and were excluded from the training data set based on the abovementioned ICME lists. $k=7$ was chosen based on the observations in \citet{heidrich2018solar}. The $7$-means classification includes two coronal hole wind types which are referred to as ($7$-means) CH1 and CH2. The source code to generate the $k$-means coronal hole wind types is available at \citet{Teichmann2023-bu}. 
\end{enumerate}

The number of data points identified as coronal hole wind differs among the four solar wind categorizations: 153594 for \citet{zhao2009global}, 74188 for \citet{xu2014new}, 33702 for $7$-means CH1, and 31070 for $7$-means CH2. Thus, while the two $7$-means coronal hole wind types together amount to a similar number of coronal hole wind observations than the \citet{xu2014new} categorization, the \citet{zhao2009global} categorization identifies approximately twice as many observations as coronal hole wind. 

Figure~\ref{fig:hist1d} provides an overview on the properties of the four considered coronal hole wind types. As an additional reference, the union of the two 7-means coronal hole wind types (referred to as $7$-means CH1+CH2) is also included. For each solar wind type and each solar wind property a 1-dimensional histogram is shown. As expected for coronal hole wind, all five types show low O charge state ratios $\nO$, low proton densities $\np$, and high proton temperatures $\Tp$. The proton-proton collisional age $\ac$ is also low as expected for the collisionless coronal hole wind (see, for example, \citet{kasper2012evolution,heidrich2020proton}). With the exception of \citet{zhao2009global} which also contains a considerable fraction of solar wind with intermediate speeds (but still mainly above $400$ km/s), the majority of coronal hole wind observations have high proton speeds $\vp$. The \citet{zhao2009global} coronal hole wind also includes lower proton temperatures and higher proton densities than the other three coronal hole wind types.

As shown in \citet{zhao2014polar,d2019slow,bale2019highly,stansby2019diagnosing,d2021alfvenic,wang2024coronal}, coronal hole wind observed from small coronal holes, in particular near the solar activity maximum, exhibit on average a lower proton temperature than coronal hole wind from equatorial coronal holes. This property is not reflected in the \citet{xu2014new} coronal hole wind which does not distinguish between subtypes of coronal hole wind. To define coronal hole wind, \citet{xu2014new} selected \textit{unperturbed coronal hole wind} from long, repeating high-speed streams\markme{. Such recurrent coronal hole wind streams are observed during the declining phase of the solar activity maximum.} This suggests that the \markme{initial} selection \markme{in \citet{xu2014new}} is probably biased toward conditions \markme{from the declining phase of the solar activity cycle ($\sim$ 2002)into the solar activity minimum ($\sim$2008) and} coronal hole wind from larger coronal holes.

The combination of the two 7-means coronal hole wind types (CH1+CH2) is very similar to the \citet{xu2014new} coronal hole wind in all seven solar wind properties. This suggests that the data-driven $7$-means identification of coronal hole wind agrees well with the \citet{xu2014new} scheme. But with $k=7$, $k$-means identified two subtypes of coronal hole wind, CH1 and CH2. Although both fit the expectations for coronal hole wind, their properties differ systematically from each other. Overall, CH1 remains more similar to the \citet{xu2014new} coronal hole wind, whereas CH2 favors (relatively) higher O and Fe charge states, a lower proton-proton collosional age, a higher magnetic field strength, and a higher proton density. Different subtypes of coronal hole wind have been identified in previous studies. For example, \citet{zhao2014polar} investigated the properties of coronal hole wind from polar and equatorial coronal holes separately and found that coronal hole wind from equatorial coronal holes tends to exhibit (slightly) higher O charge states, a lower proton-proton collisional age, and a higher magnetic field strength which fits the properties of the $7$-means CH2 type. This hints at the interpretation that $7$-means probably also separated polar from equatorial coronal hole wind. As discussed in, for example, \citet{wang2019observations}, equatorial coronal hole usually feature stronger footpoint fields which are here probably  associated with CH2, whereas weaker footpoint fields (usually observed for polar coronal holes) appear associated with CH1. \markme{Thereby, CH2 probably corresponds to Alfv{\'e}nic slow solar wind \citep{d2015origin} that originates in small low-latitude coronal holes. CH1 is more likely related to large low-latitude coronal holes that are often extensions of the polar coronal holes and tend to be found near decayed remnants of active regions and are typical sources of recurrent coronal hole wind.} This is consistent with the observations in \citet{wang2019observations,d2019slow,bale2019highly,stansby2019diagnosing,d2021alfvenic,wang2024coronal} that identify similar properties for solar wind from small equatorial coronal holes.
\citet{heidrich2016observations} reported variations in the Fe charge states even within the same coronal hole wind streams. The $7$-means classification apparently also relies on this property to characterize its second coronal hole wind type CH2.

With the possible exception of the \citet{zhao2009global} coronal hole wind, Fig.~\ref{fig:hist1d} illustrates that all four considered solar wind types indeed exhibit the properties that are expected for coronal hole wind. For the context of this study, we consider this as validation that these indeed all identify coronal hole wind.  
In Sect.~\ref{sec:comp}, we investigate these systematic differences between these four coronal hole wind types in more detail.

\section{Comparison of coronal hole wind types}
\label{sec:comp}

Already the solar wind fractions shown in Fig.~\ref{fig:fractions} illustrate considerable differences between the coronal hole wind identified by the \citet{zhao2009global} and the \citet{xu2014new} solar wind classifications. The comparison in \citet{neugebauer2016comparison} which found good agreement between the \citet{zhao2009global} and \citet{xu2014new} coronal hole winds, focused on the time period when the Genesis mission \citep{lo2001genesis} was active (late 2001 - early 2004). Fig.~\ref{fig:fractions} shows that during this time period, the coronal hole wind fractions of \citet{zhao2009global} and \citet{xu2014new} are indeed similar. However, the differences are more pronounced during the rest of the solar activity cycle. The most striking difference occurs in 2009 when the \citet{zhao2009global} scheme classifies almost all solar wind as coronal hole wind while the other three coronal hole wind types (\citet{xu2014new} coronal hole wind and $7$-means CH1 and CH2) observe only very small fractions of coronal hole wind. However, this is not the only interesting feature in Fig.~\ref{fig:fractions}. During the complete solar activity minimum period (2006-2009), \citet{zhao2009global} sees a higher fraction of coronal hole wind than the three other coronal hole wind types. This is also visible in Figure 1 of \citet{zhao2009global}. During the solar activity maximum, the observed coronal hole wind fractions are more similar (with the exception of CH1). 
It is interesting to note that from the two $7$-means coronal hole wind types, CH1 is dominant during the solar activity minimum whereas CH2 is more frequent during the solar activity maximum. Therein, during the solar activity minimum period the fraction of CH1 is similar to that of \citet{xu2014new}. But during the solar activity maximum, the CH2 fraction is similar to \citet{xu2014new}.

\subsection{Effect of changes of solar wind properties with the solar activity cycle on coronal hole wind classification}
Fig.~\ref{fig:cr} provides an overview on how solar wind properties change with the solar activity cycle (compare, for example, \citet{zurbuchen2002solar,zhao2009global,shearer2014solar}). For the sake of brevity only four solar wind parameters are shown here: $\nO$ in the top left (Panels a)-d)), $\qFe$ in the top right (Panels e)-h)), $\vp$ in the bottom left (Panels i)-l)), $\Tp$ in the bottom right (Panels m)-p)). In each panel, 1-dimensional histograms that each represent three Carrington rotations  are shown for the respective solar wind parameter. Each of the four subfigures include one panel per coronal hole wind type. 
As a reference, a red line indicates the respective median for all solar wind in each panel and red shading marks the range of all solar wind observations in each Carrington rotation from the $5$th to $95$th percentile.

On average charge states are higher during the solar activity maximum (approx 2002-2004) than during the solar activity minimum (approx 2006-2009). The effect is stronger for the O charge state ratio than for the average Fe charge state. This observation holds for each solar wind type shown here individually, and also for all solar wind observations. In particular in 2008-2009 almost all solar wind observations are below the \citet{zhao2009global} threshold (indicated with a dashed cyan line in Panels a)-d)). The trend of decreasing $\nO$ is broken in Panel a) for the \citet{zhao2009global} coronal hole wind. Here, the $\nO$ suddenly increases again. For the three other coronal hole wind types (Panels b)-d)), $\nO$ increases later with the onset of the next solar cycle in 2010. A similar (but less pronounced) effect is visible for $\qFe$ in Panels e)-h). Again, the charge states increase for the \citet{zhao2009global} coronal hole wind at least a year earlier than for the other three solar wind types. We expect that the requirement of a minimum number of counts for O and Fe in SWICS results in a selection bias which contributes to the low number of coronal hole wind observations in 2009. This requirement systematically removes in all figures some very dilute plasma, which is most likely to be observed from polar coronal holes during the solar activity minimum.

Panels i)-l) show the temporal evolution of $\vp$ from 2001-2010. The proton speed is high for all coronal hole wind types and varies from Carrington rotation to Carrington rotation. During the solar activity maximum ($\sim$2003), this variability is weakest for CH1, which tentatively implies that pure polar coronal hole wind (from coronal holes with weak footpoint strength) is rarer during this time period. The variability is strongest for the \citet{zhao2009global} coronal hole wind. Here, two tracks are visible: one with high solar wind speeds (frequently around $600$ km/s) and a second one with lower proton speeds (around and even below $400$ km/s). While proton speeds in the order of $400$ km/s are also observed for the \citet{xu2014new} coronal hole wind and the two $7$-means coronal hole wind types (CH1 and CH2), these are considerably more frequent in the \citet{zhao2009global} coronal hole wind. This apparent bifurcation in the proton speeds of the \citet{zhao2009global} coronal hole wind begins approximately in 2005 and lasts until the end of the studied time period. A similar behavior can be seen for the proton temperature in Panels m)-p)). While the \citet{xu2014new} coronal hole wind and $7$-means CH1 and CH2 show more consistently high proton temperatures during the complete part of the solar activity cycle shown here, the proton temperatures of the \citet{zhao2009global} coronal hole wind exhibit a similar bifurcation effect as for the proton speeds. From approximately 2005 to the end of 2010,  the \citet{zhao2009global} coronal hole wind contains a mixture of cold and hot plasma. In particular in the green-shaded time period in 2009 (and a few months before) only the cold and slow component remains in the \citet{zhao2009global} coronal hole wind. \citet{xu2014new} coronal hole wind (Panel n)) and CH2 (Panel p)) share a peak in the average proton temperature at the end of 2005, this could hint at strong wave activity during this time period but could also be a side-effect of constant thresholds for \cite{xu2014new} and CH2 which appear to be better adapted to solar activity maximum conditions.

These observations suggest the interpretation that the fixed-threshold of the \citet{zhao2009global} scheme misses the underlying changes in the charge state composition with the solar activity cycle and therefore likely mistakes more and more slow solar wind for coronal hole wind during the solar activity minimum. Although all solar wind categorizations considered here rely on fixed thresholds that do not change with the solar activity cycle, the \citet{zhao2009global} scheme is most affected by this. The two coronal hole wind types identified by $7$-means circumvent this problem partially insofar as CH1 represents properties of smaller coronal holes dominant during the solar activity maximum and CH2 represents properties of larger coronal holes that are dominant during the solar activity minimum.

\begin{figure}\begin{center}
  \includegraphics[width=0.9\columnwidth]{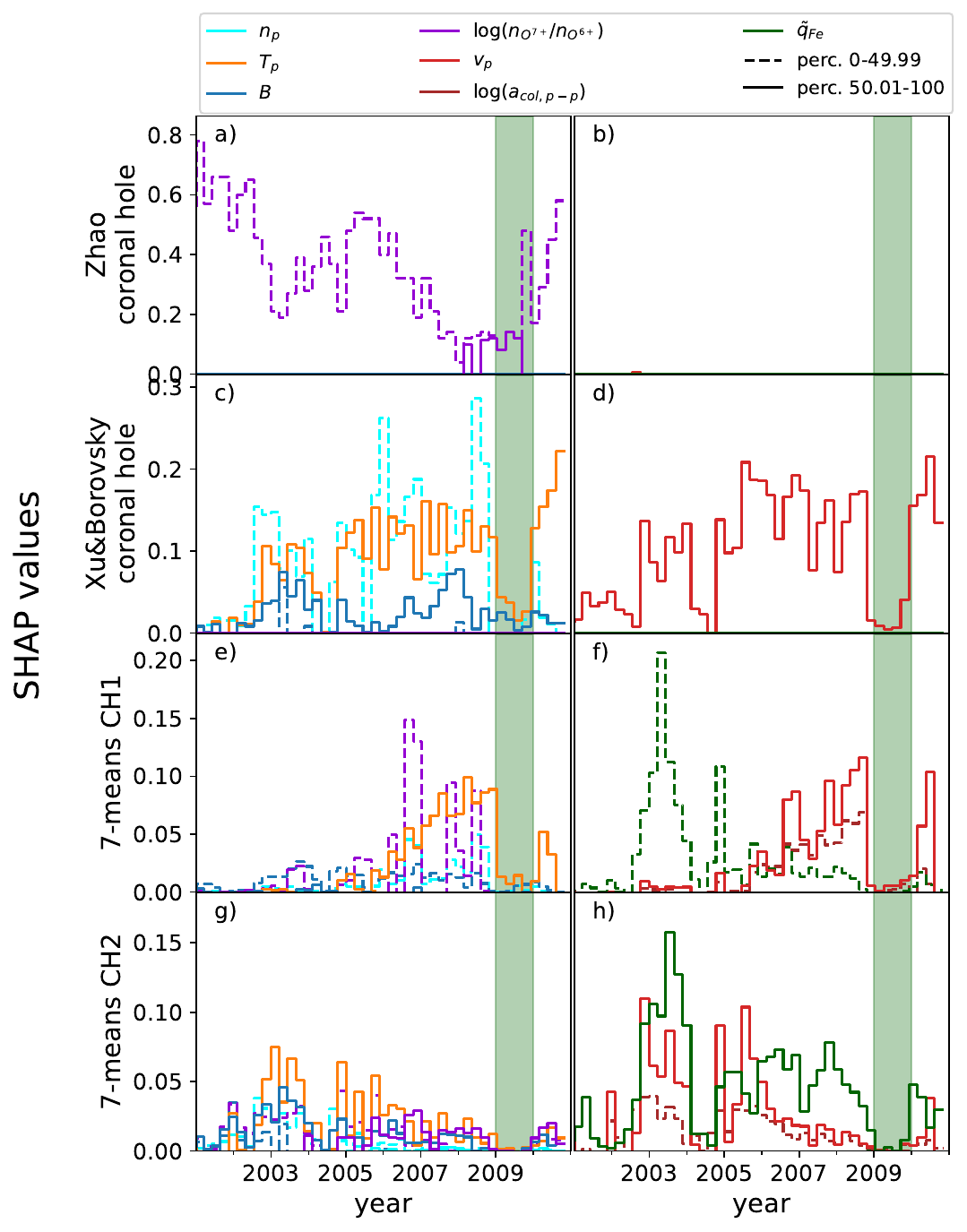}
  \end{center}
  \caption{\label{fig:shap} Averaged SHAP values per three Carrington rotations over time for solar wind parameter and coronal hole wind type. From top to bottom:  \cite{zhao2009global} coronal hole, \citet{xu2014new} coronal hole, 7-means CH1, and 7-means CH2.  Green shading highlights the year 2009. The SHAP values for seven considered solar wind parameters are distributed over the left and right columns. Left column shows: $\np$, $\Tp$, $\B$, and $\lnO$. Right column shows: $\vp$, $\ac$, and $\qFe$. For each solar wind parameter, two lines with the same color give for each solar wind parameter the importance of low parameter values (the median SHAP value of all observations below the median of all observations from 2001-2010, that is, percentiles 0-49.99, dashed lines) and high parameter values (the median SHAP value of all observations above the median of all observations from 2001-2010, that is, percentiles 50.01-100, solid lines). Only positive SHAP values are shown to focus in the importance of each parameter to identify coronal-hole wind.}
\end{figure}

\subsection{Changing importance of solar wind parameters for identifying coronal hole wind}
To provide an additional perspective, Fig.~\ref{fig:shap} shows an explainability measure that quantifies how relevant each solar wind parameter is for the identification as coronal hole wind (high absolute SHAP values $>0$). To simplify the representation, we here focus only on positive SHAP values. Small absolute values indicate that a solar wind parameter value is unimportant at this time. In addition, for each solar wind parameter Fig.~\ref{fig:shap} differentiates between high feature values (above the median, solid lines) and low feature values (below the median, dashed lines), as introduced in Sect.~\ref{sec:shap}.

Panels a) and b) give the respective SHAP values averaged over three Carrington rotations for the \citet{zhao2009global} coronal hole wind. Here, as expected, the O charge state ratio is the only parameter that is relevant for identifying coronal hole wind. The six curves for the six other parameters are included but are all plotted on top of each other at the zero line in both the left and right panels. In the right panel, a small peak for the proton speed is just visible in 2002. That the SHAP values indeed correctly identify that the parameters which are by design not relevant to identifying coronal hole wind in the \citet{zhao2009global} scheme are unimportant (that is, have low absolute SHAP values close to zero) provides a basic sanity check for the interpretation of SHAP values. Low O charge state ratios (dashed lines in Fig.~\ref{fig:shap}) are relevant to recognize solar wind as \citet{zhao2009global} coronal hole wind. Since Fig.~\ref{fig:shap} only shows positive values, the respective high importance of high O charge state ratios for identifying non-coronal hole wind is not directly visible. We omitted negative SHAP values to simplify the representation and focus only on the coronal hole types.  However, in particular during 2008-2009 the importance of the O charge state ratio for the \citet{zhao2009global} scheme is reduced and the solid line becomes visible for several Carrington rotations. This can be understood as a consequence of the following effect: during this time period (almost) all solar wind observations have $\nO$ below the decision threshold as shown in Fig.~\ref{fig:cr}. In addition, the median of all considered $\nO$ observations is with $0.095$ below the \citet{zhao2009global} threshold to identify coronal hole wind. Thus, the two curves merge, if (almost) no observations above the \citet{zhao2009global} threshold are observed during a Carrington rotation. Further, the actual value of $\nO$ does not play a role anymore (and receives a lower absolute value in its SHAP value) since the classification is here (almost) always the same. 

Panels c) and d) show the corresponding SHAP values for \citet{xu2014new} coronal hole wind. Here, most of the time high proton temperatures, low proton densities, and high proton speeds are relevant to identify coronal hole wind. Although the \citet{xu2014new} scheme also depends on the magnetic field strength, $B$ is less relevant to distinguishing coronal hole wind from non-coronal hole wind. For the \citet{xu2014new} classification, two time periods stand out with very low SHAP values: around the middle of the year 2004 and the year 2009. For these time periods, Fig.~\ref{fig:fractions} showed very low fractions of \citet{xu2014new} coronal hole wind. Thus, the precise values of the solar wind parameters are not important during these times to identify \citet{xu2014new} coronal hole wind, since (almost) all observations here are non-coronal hole wind. According to \citet{xu2014new}, coronal hole wind was characterized based on unperturbed coronal hole wind from repeating high-speed streams. Since repeating high-speed streams are more likely to be observed during the solar activity minimum, this characterization probably introduced a selection bias towards coronal hole wind from polar (or larger equatorial) coronal holes with low footpoint field strength and a corresponding bias against coronal hole wind from smaller equatorial coronal holes with higher footpoint field strength. This bias probably reduces the fraction of \citet{xu2014new} coronal hole wind in 2004 but not in 2009.

Panels e)-h) provide an overview of the importance of each considered solar wind parameter for $7$-means CH1 and $7$-means CH2. Here, all seven solar wind parameters play some role in identifying these two solar wind types. Overall,  for $7$-means CH1, low values of the O charge state ratio and the average Fe charge state (purple and dark green dashed lines in Panels e) and f)) and high values of the proton speed and proton temperature (red and orange solid lines Panels e) and f)) are the  most important (single) parameters for $7$-means CH1.  For $7$-means CH2 the average high values of the average Fe charge state (dark green solid lines in Panel h), high values of the proton speed and proton temperature (solid red and orange lines in Panels g) and h)) are most important. In agreement with Fig.~\ref{fig:fractions}, the importance of most solar wind parameters to identify $7$-means CH1 is higher during the solar activity minimum (in particular 2006-2008) and low during the solar activity maximum (in particular 2003-2004), whereas the opposite is the case for $7$-means CH2.

In 2009 the SHAP values for all four coronal hole wind types and all parameters are low because  (almost) all solar wind (as is the case for \citet{zhao2009global}) is seen as coronal hole wind, or (almost) no solar wind is identified as coronal hole wind (as is the case for the three other coronal hole wind types).


\subsection{Low fraction of coronal hole wind in 2009}
\citet{fujiki2016long} investigated the size of the coronal hole area over several solar cycles. In particular, Figures 3 and 4 in \citet{fujiki2016long} shows almost no coronal holes at low latitudes (below 30${\degree}$ latitude) during 2009. A similar effect is visible
in Figure 3 in \citet{fujiki2016long} at the ends of the previous solar cycles. In 2003, when \citet{xu2014new} sees a high fraction of coronal hole wind, Figure 4 in \citet{fujiki2016long} shows an unusually large area of equatorial coronal holes. Thus, ACE is probably connected more frequently to equatorial coronal holes in 2003 and therefore observes a high fraction of coronal hole wind.
Since ACE is located close to the ecliptic, ACE tends to be more frequently
connected to low-latitude coronal holes than to high latitude coronal
holes (in particular under stable solar activity minimum
conditions). This supports the low fraction of coronal hole wind
during 2009 determined by \citet{xu2014new} and $7$-means. The small
fraction of remaining coronal hole wind appears to originate at higher
latitudes.


\section{Conclusion}
\label{sec:con}
Coronal hole wind is considered the best-understood component of the solar wind \citep{marsch2006kinetic, cranmer2009coronal} and different solar wind categorizations have been reported to agree well in their respective coronal hole wind identification \citep{neugebauer2016comparison}. However, coronal hole wind is not uniform in its properties. Differences between equatorial and polar coronal hole wind have been identified in, for example, \citet{zhao2014polar} and \citet{wang2016role} and the O-cool coronal hole wind can be Fe-hot or Fe-cool \citep{heidrich2016observations}. Nevertheless, established solar wind categorizations that rely on fixed decision boundaries (for example, \citet{zhao2009global, xu2014new}) differ in where exactly these decision boundaries should separate coronal hole wind from slow solar wind. This study was motivated by the drastically different coronal hole wind fractions observed by \citet{zhao2009global} and \citet{xu2014new} during the solar activity minimum. As an alternative, purely data-driven perspectives coronal hole wind identified by an unsupervised machine learning method $k$-means (here, with $k=7$) was included in the comparison. Therein, $7$-means identifies two types of coronal hole wind, here called CH1 and CH2. Both exhibit typical properties of coronal hole wind, but while CH1 (based on the properties of polar and equatorial or weak and strong footpoint-field coronal hole wind reported in \citet{zhao2014polar,wang2016role,wang2019observations}) appears to favor polar coronal hole wind with weak footpoint fields, CH2 appears to mainly include equatorial coronal hole wind with high  average Fe charge state and high footpoint strengths. 

We identified the reliance on fixed thresholds as the main cause for the disparity in the determined coronal hole fractions of \citet{zhao2009global} compared to the other three coronal hole wind types: The O charge state ratio $\nO$ is known to vary systematically with the solar activity cycle \citep{zurbuchen2002solar,zhao2009global,shearer2014solar}. The fixed threshold in \cite{zhao2009global} appears to be best adjusted to solar activity maximum conditions when the agreement with \citet{xu2014new} is high as discussed in \citet{neugebauer2016comparison}, but appears less appropriate for the solar activity minimum. For the \citet{zhao2009global} scheme this evidently leads to a misclassification of slow solar wind as coronal hole wind during the solar activity minimum. The \citet{xu2014new} coronal hole wind appears to be better adjusted to solar activity minimum than solar activity maximum conditions.

The very low fraction of coronal hole wind identified in 2009 by \citet{xu2014new} and $7$-means could also be a result of their respective fixed decision boundaries since these methods are also adversely affected by ignoring the systematic changes in the solar wind with the solar activity cycle. However, this low fraction of coronal hole wind observed at L1 by ACE in 2009 is probably also related to the very small number of coronal holes at low latitudes in 2009 \citep{fujiki2016long}. 

We conclude that solar wind categorization needs to be revisited even to identify coronal hole wind. Such improved solar wind categorizations should take the phase of the solar cycle explicitly into account. Further, an optimal set of input parameters for solar wind classification should be refined and include a more direct measure of wave-activity, for example, the Alfv{\'e}nicity parameter \citet{ko2018boundary,wang2019observations}.

\section*{Acknowledgements}
  This work was supported by the \emph{Deutsches Zentrum für Luft-
      und Raumfahrt} (DLR) as SOHO/CELIAS 50 OC 2104.  We further
      thank the science teams of  ACE/SWEPAM, ACE/MAG as well as
      ACE/SWICS for providing the respective level 2 and level 1 data
      products.
%
   \bibliographystyle{arxiv} 
   \bibliography{aa} 
%

\end{document}